\documentclass[aps,prb,showpacs,groupedaddress,twocolumn]{revtex4}
\usepackage{amsmath,amssymb}
\usepackage{graphicx} 
\usepackage{dcolumn} 
\usepackage{bm} 

\begin{document}
\draft
\title{\bf Transient dynamics of molecular devices under step-like pulse bias}

\author{Bin Wang, Yanxia Xing, Lei Zhang, and Jian Wang$^*$}

\address{
Department of Physics and the Center of Theoretical and
Computational Physics, The University of Hong Kong, Pokfulam Road,
Hong Kong, China}

\begin{abstract}
We report first principles investigation of time-dependent current
of molecular devices under a step-like pulse.
Our results show that
although the switch-on time of the molecular device is comparable to
the transit time, much longer time is needed to reach the steady
state. In reaching the steady state the current is dominated by
resonant states below Fermi level. The contribution of each resonant
state to the current shows the damped oscillatory behavior with
frequency equal to the bias of the step-like pulse and decay rate
determined by the life time of the corresponding resonant state. We
found that all the resonant states below Fermi level have to be
included for accurate results.
This indicates that going beyond wideband limit is essential for a
quantitative analysis of transient dynamics of molecular devices.
\end{abstract}

 \maketitle

Anticipating a variety of technological applications, molecular
scale conductors and devices are the subject of increasingly more
research in recently years. One of the most important issues of
molecular electronics concerns the dynamic response of molecular
devices to external
parameters\cite{gross3,Zhu,yip,Maciejko,diventra,chen1,gross2}. For
ac quantum transport in such small devices, atomic details and
non-equilibrium physics must be taken into account. So, in
principle, one should use the theory of non-equilibrium Green's
function (NEGF)\cite{Jauho2} coupled with the time-dependent density
functional theory (TDDFT)\cite{gross1} to study the time-dependent
transport of molecular devices. Practically, it is very difficult to
implement it at present stage due to the huge computational cost.
One possible way to overcome this problem is to use the adiabatic
approximation, an approach widely used in mesoscopic physics. In
this approach, one starts from a steady-state Hamiltonian and adds
the time dependent electric field adiabatically. This is a
reasonable approximation since most of the time the applied electric
field is much smaller than the electrostatic field inside the
scattering region. In addition, it has been shown
numerically\cite{chen1} that dc transport properties such as I-V
curve obtained from the equation of motion method coupled with TDDFT
agrees with that obtained by the method of NEGF coupled with the
density functional theory (DFT)\cite{mcdcal,brand}. Hence, under the
adiabatic approximation, one could replace TDDFT by DFT and use the
NEGF+DFT scheme to calculate ac transport properties of molecular
devices.

We consider a system that consists of a scattering region coupled to
two leads with the external time dependent pulse bias potential
$v_\alpha(t)$. The time-dependent current for a step-like pulse has
been derived exactly going beyond the wide-band limit by Maciejko et
al\cite{Maciejko}. Since the general expression for the current
involves triple integrations, it is extremely difficult if not
impossible to calculate the time-dependent current for real systems
like molecular devices. In this regard, approximation has to be made
in order to carry out time-dependent simulations of molecular
devices. We note that the simplest approximation is the so called
wide-band approximation where self-energies $\Sigma^{r,a}$ are
assumed to be independent of energy.\cite{Jauho1} Indeed, if such an
approximation is used, i.e., $\Sigma^r=-i\Gamma/2$, one recovers the
expression of transient current first obtained by Wingreen et
al\cite{Jauho1}. However, there are two problems when applying this
approximation to investigate the dynamics of molecular devices.
First of all, one assumes implicitly that the contribution to the
transient current is dominated by only one resonant level with a
constant linewidth function $\Gamma$ in the system in such an
approximation. As we shall show below that this is not a good
assumption in first principles investigation of the dynamics of
molecular devices because there could be several resonant levels
that significantly contribute to the transient current in molecular
devices. Secondly, in the steady state limits at $t=0$ and $=\infty$
the wide-band limit can not reproduce the correct dc I-V curve
obtained from first principles.
By assuming the wide-band limit one can get a very different current
that depends on the choice of $\Gamma$. In this paper, we propose an
approximate formula of transient current that is suitable for
numerical calculation for real molecular devices. Our scheme is an
approximation of the exact solution of Maciejko et al\cite{Maciejko}
while keeping essential physics of dynamic systems. Using this
scheme, we have calculated the transient current for several
molecular devices. We found that all the resonant states below Fermi
level contribute to the transient current. Each resonant state gives
a damped oscillatory behavior with frequency equal to the bias of
pulse and decay rate equal to its life time. Because of sharp
resonances, it takes much longer time for the current to relax to
the equilibrium value. For instance, for a $Al-C_4-Al$ structure
with a transit time of $L/v_F=1.3fs$ the relaxation time is about
50fs. For a CNT-DTB-CNT structure with a transit time of 1fs,
however,  the relaxation time can reach several ps due to the
resonant state with long lifetime. Our results indicated that going
beyond wide-band limit is crucial for accurate predictions of
dynamic response of molecular devices.

From Ref.\onlinecite{Jauho1,Maciejko}, the current is expressed as
($\hbar=q=1$),

\begin{eqnarray}
J_\alpha(t)=2{\rm Re}\int \frac{d\epsilon}{2\pi} {\rm Tr}[{\cal
J}_\alpha(t,\epsilon)] \label{eq1}
\end{eqnarray}
where
\begin{eqnarray}
{\cal J}_\alpha(t,\epsilon) =
A_\alpha(t,\epsilon)\Sigma^{<,0}_\alpha(\epsilon) + \sum_\beta
A_\beta(t,\epsilon)\Sigma^{<,0}_\beta(\epsilon)
F_{ \beta\alpha}(t,\epsilon)
\label{jout}
\end{eqnarray}
where $\Sigma^{<,0}$ and $\Sigma^{a,0}$ are equilibrium
self-energies and $A_\alpha(t,\epsilon)$ and $F_{
\beta\alpha}(t,\epsilon)$ have different definitions for upward and
downward pulses (see Ref.\onlinecite{Maciejko} for details). In the
absence of ac bias, $A_\alpha(t,\epsilon)$ is just the Fourier
transform of retarded Green's function. As discussed in
Ref.\onlinecite{Jauho2} that the first term in Eq.(\ref{jout})
corresponds to the current flowing into the central scattering
region from lead $\alpha$ while the second term corresponds to the
current flowing out from the central region into lead $\alpha$. From
Eq.(\ref{eq1}) we see that in order to calculate the transient
current for a pulse bias we need to include the states with energy
from $-\infty$ to the Fermi energy. This is very different from dc
case where only the states with energy in the range $v_L-v_R$ about
Fermi level contribute. Physically, this can be understood as
follows. For ac transport with a sinusoidal bias $\cos(\omega t)$,
the photon assisted tunnelling is significant only for the first a
few sidebands\cite{Jauho2}. The step-like pulse can be expanded in
terms of sinusoidal bias with continuous distribution of frequencies
and each sinusoidal bias generates a photon sideband that
facilitates the photon assisted tunnelling. Hence we expect that all
the resonant states below Fermi level should be included and
carefully examined in the calculation of transient current. Note
that Eq.(\ref{eq1}) and (\ref{jout}) are exact expressions with
$A_\alpha(t,\epsilon)$ and $F_{ \beta\alpha}(t,\epsilon)$ given in
Ref.\onlinecite{Maciejko}. Our approximation is made on
$A_\alpha(t,\epsilon)$ and $F_{ \beta\alpha}(t,\epsilon)$. For the
upward pulse, $A_\alpha(t,\epsilon)$ and $F_{
\beta\alpha}(t,\epsilon)$ are given by the following ansatz,
\begin{eqnarray}
A^u_\alpha(t,\epsilon) =
A^u_{1\alpha}(t,\epsilon)+A^u_{2\alpha}(t,\epsilon) \label{aa}
\end{eqnarray}
with
\begin{eqnarray}
A^u_{1\alpha}(t,\epsilon)&=&\int \frac{dE}{2\pi i}
\frac{e^{i(\epsilon-E+v_\alpha)t}}{E-\epsilon-i0^+}~{\bar
G}^{r}_0(E,\epsilon) \nonumber
\\
A^u_{2\alpha}(t,\epsilon)&=& \int \frac{dE}{2\pi
i}\frac{1-e^{i(\epsilon-E+v_\alpha)t}}{E-\epsilon-v_\alpha-i0^+}~{\bar
G}^{r}_\alpha(E,\epsilon)  \label{Au}
\end{eqnarray}
and
\begin{eqnarray}
[F^u_{\beta\alpha}(t,\epsilon)]^\dagger =
\Sigma^{r,0}_\alpha(\epsilon)A^u_{1\beta}
+\Sigma^{r,0}_\alpha(\epsilon-v_\alpha+v_\beta)A^u_{2\beta}
\label{Fu}
\end{eqnarray}
where
\begin{eqnarray}
{\bar G}^{r}_0(E,\epsilon) =
1/[E-H-U_{eq}-\Sigma^{r,0}_{~}(\epsilon)]\label{Gr0}
\end{eqnarray}
\begin{eqnarray}
{\bar G}^{r}_\alpha(E,\epsilon) = 1/[E-H-U-\sum_\beta
\Sigma^{r,0}_\beta(\epsilon+v_\alpha-v_\beta)]\label{GrR}
\end{eqnarray}
with $U_{eq}$ and $U$ are, respectively, the equilibrium Coulomb
potential and dc Coulomb potential at bias $v_L-v_R$.
As will be illustrated in the examples given below this ansatz can
be easily implemented to calculate the transient current for real
molecular devices. Importantly, the results obtained from the ansatz
captured essential physics of molecular devices. We wish to
emphasize that our ansatz goes beyond the wide-band limit. It agrees
with the expression of time-dependent current obtained by Wingreen
et al in the wide-band limit\cite{Jauho1} and produces correct
limits at $t=0$ and $t \rightarrow \infty$.

Note that ${\bar G}^{r}_0(E,\epsilon)$ and ${\bar
G}^{r}_\alpha(E,\epsilon)$ are different from the usual definition
of Green's functions, they allow us to perform contour integration
over energy $E$ in Eq.(\ref{Au}) and (\ref{Fu}) by closing a
contour with an infinite radius semicircle at lower half plane.
For a constant $\epsilon$, we have the following eigen equations
\begin{eqnarray}
\left(H+U_{eq}+\Sigma^{r,0}(\epsilon)\right)|\psi^0_n\rangle
&=&\epsilon^0_n |\psi^0_n\rangle  \nonumber \\
\left(H+U_{eq}+\Sigma^{a,0}(\epsilon)\right)|\varphi^0_n\rangle
&=&\epsilon^{0*}_n |\varphi^0_n\rangle. \label{eigen1}
\end{eqnarray}
Expanding ${\bar G}^{r}_0(E,\epsilon)$ in terms of its eigen
functions $|\psi^0_n\rangle$ and $|\phi^0_n\rangle$, we
have\cite{foot1}
\begin{eqnarray}
{\bar G}^{r}_0(E,\epsilon) = \sum_n |\psi^0_n\rangle
\langle\phi^0_n|/(E-\epsilon^0_n+i0^+)\label{expand}.
\end{eqnarray}
With similar expression for ${\bar G}^{r}_\alpha(E,\epsilon)$,
Eq.(\ref{Au}) can be written as\cite{foot10}
\begin{eqnarray}
A^u_{1\alpha} &=& \sum_n
\frac{e^{i(\epsilon-\epsilon^0_n+v_\alpha)t}}{\epsilon-\epsilon^0_n+i0^+}|\psi^0_n\rangle
\langle\phi^0_n|
 \nonumber \\
A^u_{2\alpha} &=& \sum_n
\frac{1-e^{i(\epsilon-\epsilon_n+v_\alpha)t}}{\epsilon-\epsilon_n+v_\alpha+i0^+}
 |\psi_n\rangle
\langle\phi_n| \label{Au1}.
\end{eqnarray}

Now we show that our formalism gives the correct limits. At $t=0$
we have $A^u_\alpha(t,\epsilon)=G^r_0(\epsilon)$ and
$F^u_{\beta\alpha }(t,\epsilon)=G^a_0(\epsilon)
\Sigma^{a,0}_\alpha(\epsilon)$ with $G^r_0(\epsilon)$ the
equilibrium Green's function. This shows that the current from
Eq.(\ref{jout}) is zero. Since all the poles $\epsilon^0_n$ and
$\epsilon_n$ in Eq.(\ref{Au1}) are on the lower half plane, at $t
\rightarrow \infty$ we have $A^u_\alpha=G^r(\epsilon+v_\alpha)$
and $F^u_{\beta\alpha
}=G^a(\epsilon+v_\beta)\Sigma^{a,0}_\alpha(\epsilon+v_\beta-v_\alpha)$
where $G^r(\epsilon)$ is the Green's function with dc bias
$v_\alpha$ at $t \rightarrow \infty$. Substituting expressions of
$A^u_\alpha$ and $F^u_{\beta\alpha }$ into Eq.(\ref{eq1}), it
gives the same dc current at the bias $v_L-v_R$. So far, we have
discussed the ac {\it conduction} current $J_\alpha(t)$ under
pulse-like bias. The displacement current $J_\alpha^d$ due to the
charge pileup $dQ/dt$ inside the scattering region can be included
using the method of current partition\cite{buttiker4,wbg}:
$J^d_\alpha=-(J_L+J_R)/2$, so that the the total current is given
by $I_L = (J_L-J_R)/2$.\cite{Jauho2}

With the formalism established, we now proceed to calculate the
dynamic response of molecular devices. We have used the first
principle quantum transport package ${\rm MatDcal}$.\cite{mcdcal}
To calculate the transient current for step-like pulse, we need to
go through the following steps: (1). calculate two potential
landscapes using NEGF-DFT package: the equilibrium potential
$U_{eq}$ at $t=0$ and the dc potential $U$ at $t=\infty$. (2).
With $U_{eq}$ and $U$ obtained, one solves eigenvalue problem
using Eq.(\ref{eigen1}) and its counterpart for $U$, then find
$A^u_{1\alpha}$ and $A^u_{2\alpha}$ from Eq.(\ref{Au1}), and
finally $A^u_\alpha$ and $F^u_{\beta \alpha}$ can be calculated
from Eq.(\ref{aa}) and (\ref{Fu}).

\begin{figure}
\includegraphics[width=9cm,height=6cm]{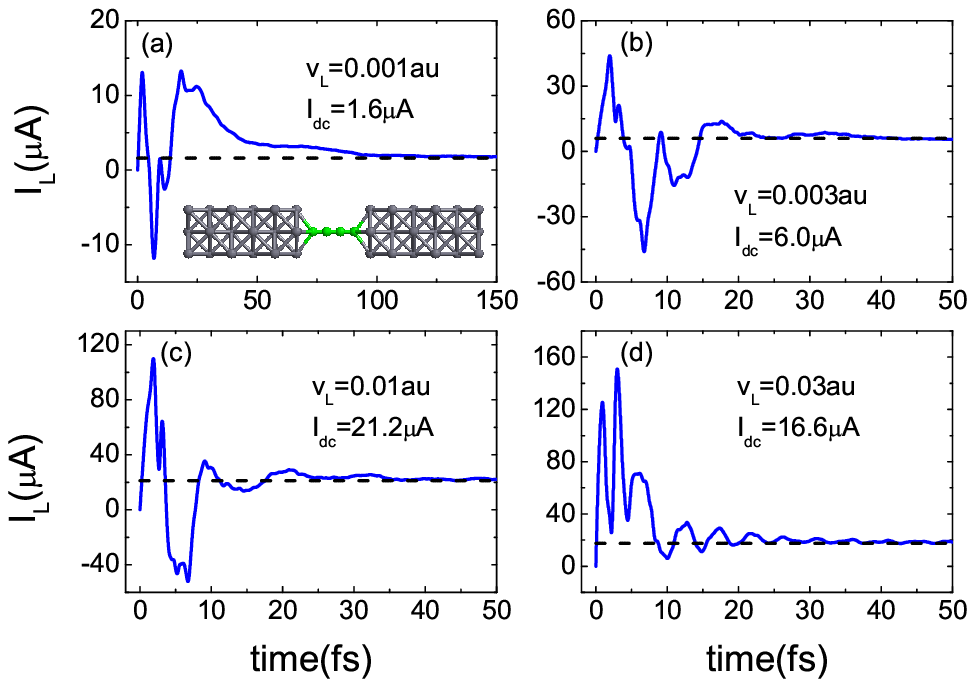}
\caption{(color online) Transient current of $Al-C_{4}-Al$ structure
at different bias $v_L=0.001, 0.003, 0.01, 0.03au$. The solid line
shows the transient current and the dotted line is the dc current
$I_{dc}$ at bias $v_L-v_R=2v_L$. Inset of Fig.1a: schematic plot of
the $Al-C_{4}-Al$ structure. } \label{fig1}
\end{figure}


Inset of Fig.1a shows the structure of $Al-C_{4}-Al$ where $Al$
leads are along (100) direction.
The nearest distance between $Al$ leads and the carbon chain is
3.781$au$ and the distance of C-C bond is 2.5$au$.($1au = 0.529
\AA$). Fig.1 shows the total transient currents $I_L(t)$ of the
$Al-C_4-Al$ structure with various voltages $v_L(t)=-v_R(t)$.
Following observations are in order. First of all, for all bias
voltages the transient currents reached the correct limits at $t=0$
and $t=\infty$. Secondly, we see that once step-like voltage is
turned on in the lead, currents oscillate rapidly with large
amplitude in the first a few fs and then gradually approach to the
steady-state values ($I_{dc}$ shown in the figure). In the first 10
to 30 fs, the current is much larger than that of the steady state
value which agrees with the results obtained using first principle
calculation with wide-band limit\cite{chen1}. For Fig.1a, the
relaxation time (time to reach to steady state) is roughly 150 fs
and for Fig.1b-1d the relaxation time is about 50fs. In addition,
the switch-on time (the time to reach the maximum current) for the
$Al-C_4-Al$ structure does not depend on the bias voltages. The
typical switch-on time is about 2fs for applied bias voltage $v_L$
ranging from $0.001au$ to $0.01au$ $(1au=27.2V)$. For Al leads, the
Fermi velocity is about $2 \times 10^{6}$ m/s which corresponds to a
transit time of 1.3 fs for the $Al-C_4-Al$ structure whose size is
about $L=47$ $au$. Thirdly, we observe that the dc limit $I_{dc}$ at
$v_L=0.01au$ is larger than that at $v_L=0.03au$. This is due to the
appearance of the negative differential resistance at about $v_L =
0.0075au$. Finally, there are several timescales characterizing the
dynamic response of the molecular device. This can be seen clearly
from Fig.1d that after 10 fs, the system shows a damped oscillation
similar to the charging process of a classical RLC circuit. We will
discuss this kind of oscillation in detail in the second example.

\begin{figure}
\includegraphics[width=9cm,height=6cm]{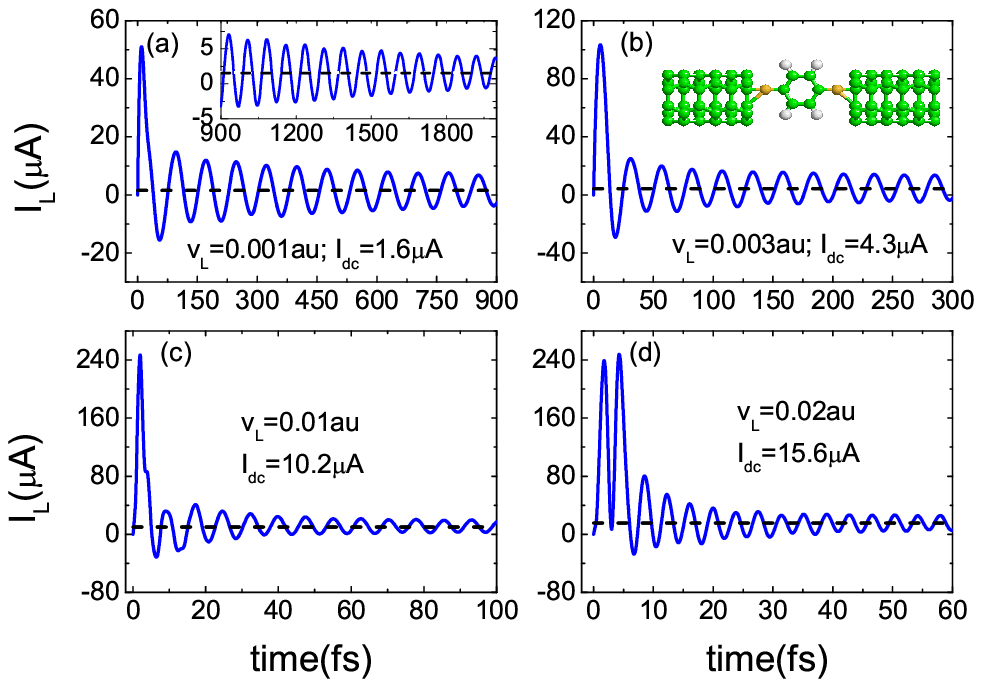}
\caption{(color online) Transient current of CNT-DTB-CNT structure
at different bias $v_L=0.001, 0.003, 0.01, 0.02au$. The solid line
shows the transient current and the dotted line is the dc current
at bias $v_L-v_R=2v_L$. Inset of Fig.2b: schematic plot of the
CNT-Di-thiol benzene-CNT structure.} \label{fig2}
\end{figure}

As a second example, we study the transient current for
di-thiol-benzene molecule (DTB) in contact with two (3,3) carbon
nanotube (CNT) leads (see inset of Fig.2b). The structure is relaxed
with the distance between the S atom and the nearest C atom equal to
2.73$au$ and the bond length of C-C being 3.61$au$. Fig.2 shows the
transient current for different upward pulse biases. We see that for
small bias $v_L=0.001au$, the current drops quickly in first 50 fs
and then oscillates with much slower decay rate. It is found that
the oscillatory part of the transient current is dominated by
$\cos(2 v_L t)$ which remains valid for the transient current at
other biases $v_L$ shown in Fig.2b to Fig.2d. For instance, this
gives the distance between adjacent peaks $\tau_0=76 fs$ in Fig.2a
when $v_L=0.001au$. With $1au=0.00242fs$, we obtain
$\tau_0=\pi/v_L$. Different from the $Al-C_{4}-Al$ structure, it
takes much longer time for the system to reach the equilibrium
current $I_L(\infty)=1.5\mu A$ (after 5000fs the current is about
$1.7\mu A$). From Fig.2a-2d, we conclude that the relaxation time is
several ps.

Physically, this can be understood from the transmission coefficient
$T(E)$. Fig.3 depicts $T(E)$ vs energy ranging from the transmission
threshold to Fermi energy. We have scanned 100,000 energy points in
order to resolve sharp resonant peaks labelled in Fig.3. In our
calculation, 100 energy points were used for each sharp resonant
peak (total 3000 energy points used) to converge the integration
over $\epsilon$, i.e., $\int d\epsilon {\rm Tr} [{\cal
J}_\alpha(t,\epsilon)]$ in Eq.(1). Since these sharp resonant peaks
correspond to resonant states with large lifetimes, the incoming
electron can dwell for a long time at these resonant states and
hence the corresponding current decays much slower than the other
states. If we focus on a particular resonant state with resonant
energy $\epsilon_0$ (below Fermi level) and half-width $\Gamma_0$,
then Eq.(\ref{aa}) gives $A^u_\alpha \sim
\exp[i(\epsilon-\epsilon_0+v_\alpha)t-(\Gamma_0/2) t]$\cite{Jauho1}.
Assuming that the sharp resonant state gives major contribution to
the current (the wideband approximation), we have $A^u_\alpha \sim
\exp(i v_\alpha t-(\Gamma_0/2) t)$. Therefore the first term in
Eq.(\ref{jout}) exhibits an oscillatory part $\exp(i v_\alpha
t-(\Gamma_0/2) t)$ while the second term behaves like $\exp(2i
v_\alpha t-\Gamma_0 t)$. It is the interplay between these two terms
that gives rise to the transient current. For instance, for Fig.2
the second term $\exp(2i v_\alpha t-\Gamma_0 t)$ dominates while for
Fig.1d the first term gives the most contribution.

Indeed, our numerical result confirms this analysis. It shows that
these resonant peaks give major contributions to the transient
current for $t>50fs$. In addition, we find that there is an one to
one correspondence between the resonant peak at $\epsilon$ and the
corresponding ${\rm Tr}[{\cal J}_\alpha(t,\epsilon)]$: ${\rm
Tr}[{\cal J}_\alpha(t,\epsilon)]$ exhibits a huge peak whenever
$\epsilon$ is near the resonance. This correspondence is important
because it indicates that our ansatz has kept essential physics
arising from the above analysis. Furthermore, our result shows that
the transient current due to each resonant peak has the same
characteristics frequencies $v_L$ or $2v_L$.

\begin{figure}
\includegraphics[width=9cm,height=6.cm]{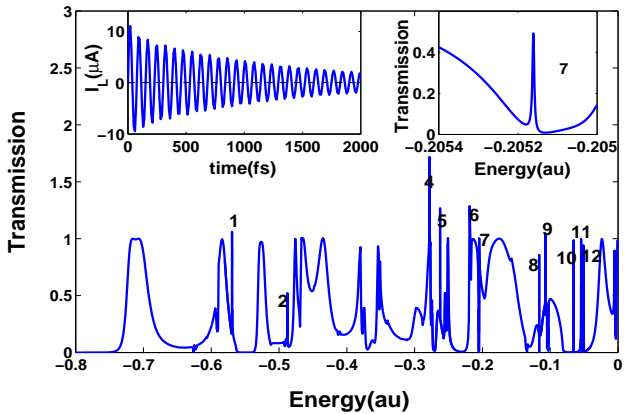}
\caption{(color online) The transmission coefficient for the
CNT-DTB-CNT structure. The insets are $T(E)$ vs energy at peak 7
and its corresponding current.} \label{fig3}
\end{figure}

Let's examine the contribution of each resonant state to the
oscillatory part of the transient current at $v_L=0.001au$ (Fig.2a).
Among these resonant peaks in Fig.3, the most contribution comes
from the peak number 7 with half-width $\Gamma_7 = 2.7 \times
10^{-5} au$ which corresponds to a decay time $\tau_7 = 1240fs$ from
the expression $\exp(-\Gamma_7 t)$. In the left inset of Fig.3, we
plot the current obtained by integrating ${\rm Tr}[{\cal
J}_\alpha(\epsilon)]$ over the neighborhood of peak 7 (see right
inset of Fig.3). It shows that the decay time is indeed
characterized by $\tau_7$. Comparing Fig.2 and Fig.3, we see that
the contribution from the peak 7 to the total current is about
$15\%$ for $t<50 fs$ while for $t>50 fs$ the contribution is $50\%$.
The next dominant contribution is due to the peaks numbered 5, 10,
and 12 whose contributions are one order of magnitude smaller. This
indicates that one has to include all the resonant peaks for
accurate results. Since different resonant peak corresponds to a
different half-width $\Gamma$, one can not choose just one $\Gamma$
to characterize the system. We have also calculated the transient
current for the structure of $Al-C_{60}-Al$ and our results show
that the long time behavior is dominated by two resonant peaks with
different $\Gamma$ and shows beat pattern with relaxation time about
800fs.

In summary, we have carried out first principles investigation of
time response of molecular devices. We found that the resonant
states below Fermi level are crucial for time-dependent
quantitative analysis. Our results indicated that the long time
behavior of transient current is dominated by resonant states and
the individual resonant state gives the damped oscillatory
behavior with frequency equal to the bias of pulse and decay rate
equal to the life time of the corresponding resonant state. Our
results indicated that one has to go beyond the wide-band limit
for quantitative calculations of dynamic response of molecular
devices.

{\bf Acknowledgments:} This work was supported by a RGC grant (HKU
704308P) from the government of HKSAR.


\begin{thebibliography}{00}
\bibitem{foot1}
In ab initio calculation, if atomic orbits are chosen, their overlap
matrice $S$ have to be included in all Green's functions by
replacing $E$ with $E S$. In this case Eq.(\ref{expand}) has to be
divided by a factor $<\phi^0_n|S|\psi^0_n>$.
\bibitem{foot10}
In order to use this formalism (Eq.(\ref{Au1})) correctly, one
must be careful in the self-energy calculation. In the numerical
calculation of self-energy, one usually adds a small imaginary
part to the real energy $E \rightarrow E+i\eta$ in order to
resolve the retarded or advanced self-energies. Unfortunately,
this parameter $\eta$ will introduce many spurious poles in the
lower half plane with imarginary part of the poles less than
$\eta$. This has no effect on ac current if it is calculated
directly. However, if the theorem of residue is used such as
Eq.(\ref{Au1}) the transient behavior will be dominated by the
spurious poles. To eliminate this effect, one has to calculate the
self-energy by setting $\eta=0$.
\end{thebibliography}
\end{document}